\documentstyle[prabib,aps,epsf]{revtex}
\begin{document}
\draft
\title{
On the Coulomb interaction in chiral-invariant 
           one-dimensional electron systems}
\author{S. Bellucci $^1$ and  J. Gonz\'alez $^2$  \\}
\address{
        $^1$Laboratori Nazionali di Frascati.
        INFN.
        P. O. Box 13. I-00044 Frascati. Italy. \\
        $^2$Instituto de Estructura de la Materia.
        Consejo Superior de Investigaciones Cient{\'\i}ficas.
        Serrano 123, 28006 Madrid. Spain. }
\date{\today}
\maketitle
\begin{abstract}

We consider a one-dimensional electron system, suitable for the
description of the electronic correlations in a metallic carbon nanotube.
Renormalization group methods are used to study the low-energy
behavior of the unscreened Coulomb interaction
between currents of well-defined chirality.
In the limit of a very large number $n$ of subbands we find
a strong renormalization of the Fermi velocity, reminiscent
of a similar phenomenon in the graphite sheet. For small $n$
or sufficiently low energy, the Luttinger liquid behavior
takes over, with a strong wavefunction renormalization leading
to a vanishing quasiparticle weight. Our approach is appropriate
to study the crossover from two-dimensional to one-dimensional
behavior in carbon nanotubes of large radius.

\end{abstract}

The recent experimental availability of single-walled fullerene
nanotubes has renewed the interest in the study of
one-dimensional electron systems\cite{nano,tans,wild,odom,kim}.  
In one spatial dimension the
Luttinger liquid concept replaces the Fermi liquid picture, and
provides the paradigm of a system with strong electronic
correlations\cite{sol,hal,vt,gz}. 
The investigation of the generic electronic instabilities of 
metallic nanotubes has been accomplished in Refs.
\onlinecite{balents} and \onlinecite{louie}.
There have been also recent attempts to look for signatures
of Luttinger liquid behavior 
in single-walled nanotubes that are
packed in the form of ropes\cite{smalley}.
Another interesting instance
that seems to be feasible from the experimental point of view
is that of nanotubes in the absence of external screening 
charges or, at least, with a 
screening length much larger than the typical transverse dimension.
The phenomenology of these systems has been studied in
Refs. \onlinecite{fisher,eg,odintsov}. Anyhow, as long as
the description of the Luttinger liquid behavior is
usually made under the assumption of a short-range
interaction, the carbon nanotubes with unscreened Coulomb
interaction should deserve further theoretical analysis, devoted
to ascertain possible deviations from the standard
Luttinger liquid picture.

\begin{figure}

\par
\centering
\epsfbox[0 700 331 1031]{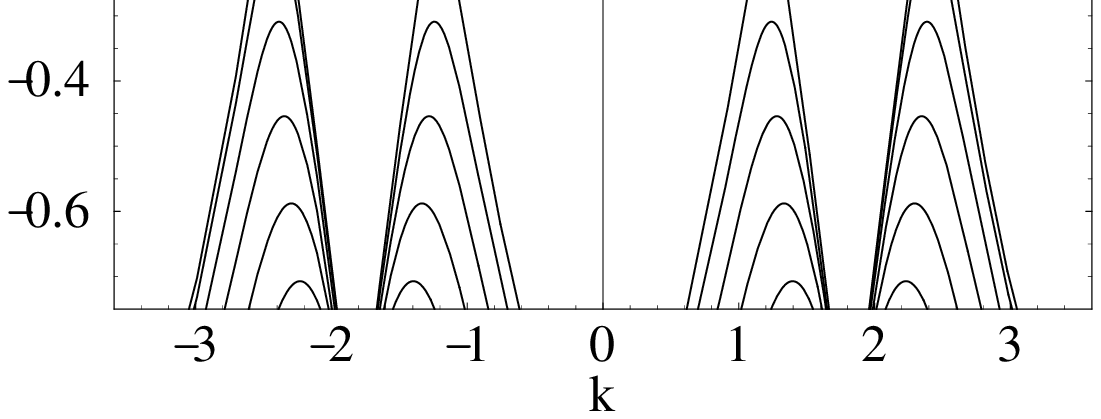}
\par

\vspace{2cm}
\caption{Low-energy band structure of a carbon nanotube with $n = 20$
subbands. The energy is measured in units of the hopping parameter and 
the momentum is in units of the inverse lattice spacing.}
\label{one}
\end{figure}

In the particular case of the metallic carbon nanotubes
\cite{class}, the Coulomb
interaction is also special in that the long-range potential
does not lead to the hybridization between left- and
right-moving electrons in the nanotube. At half-filling, the
metallic nanotubes have two Fermi points, characterized
respectively by the large momenta $K_F$ and $-K_F$, with the
typical band structure shown in Fig. \ref{one}. The underlying
lattice is such that it allows to arrange the modes of the two
linear branches at each Fermi point into a Dirac-like spinor,
whose components stand for the respective amplitudes in the two
sublattices of the honeycomb lattice\cite{ggv}. The kinetic part of the
hamiltonian can be approximated as
$H_{kin} = -t a\: \delta k_x \: \sigma_1 $, where $a$ is the
lattice spacing and $t$ is the hopping parameter.
This means that the right-movers have an amplitude $\Psi ({\bf r})$ 
that alternates sign from one sublattice to the other, while the
left-movers keep the same sign on both of them. When projecting
the Coulomb interaction to the longitudinal dimension of the tube,
one has to sum over the points of the two transversal rings
of the tube ${\cal C}_x$ and ${\cal C}_{x'}$

\begin{equation}
\langle \Psi_{\alpha }(x) \Psi_{\beta }(x') 
 \left| V(x - x') \right|
 \Psi_{\gamma }(x) \Psi_{\delta }(x') \rangle =
\sum_{i \in {\cal C}_{x}, j \in {\cal C}_{x'} }
 \Psi^{+}_{\alpha }({\bf r}_i) \Psi_{\gamma }({\bf r}_i) 
  \frac{1}{\left| {\bf r}_i - {\bf r}_j' \right| }
 \Psi^{+}_{\beta }({\bf r}_j') \Psi_{\delta }({\bf r}_j') 
\end{equation}
It becomes clear that, for a distance $x - x'$ much greater than
the lattice spacing, the matrix element is only nonvanishing
when $\alpha $ and $\gamma $, as well as $\beta $ and $\delta $,
have the same (left or right) chirality.

Thus, the relevant interaction in the study of the metallic carbon
nanotubes turns out to be the long-range Coulomb interaction
between currents of well-defined chirality. 
In this work we study
the genuine effects due to the $1/|x|$ interaction in this kind
of one-dimensional systems, focusing
on the physics at each of the Fermi points.
We disregard in this way backscattering processes and
residual short-range interactions mixing chiralities, as they have
a smaller nominal strength ($\sim 0.1 \; e^2/n$, in terms of the number
$n$ of subbands\cite{fisher,eg})
and they stay small down to extremely low
energies ($\sim t \; \exp (-75) $ \cite{eg}).
Our present purpose is to discern the low-energy properties 
of the system and, in particular, whether the long-range interaction
remains unscreened to arbitrarily large distances. It is known
that the bosonization approach yields divergent results for some
of the observables like, for instance, the plasmon velocity at 
vanishing momentum. This shortcoming is remedied in practice
by introducing some infrared cutoff dictated by the external
environment of the one-dimensional system. Anyhow, it is worthwhile to 
analyze to what extent the singular interaction may be 
renormalized in the infrared by means of a dynamical
screening effect. This problem may be also relevant when studying
the properties of nanotubes of large radius, since in those
systems a crossover from two-dimensional to one-dimensional
behavior is to be expected when taking the infrared limit.
It is known that the Coulomb interaction is strongly renormalized
on the graphite sheet\cite{ggv2}, what points again at the question about
the nature of the screening effects in carbon nanotubes with
large number of subbands.

In this Letter we use renormalization group (RG)
methods to find the low-energy effective theory of the $1/|x|$ 
interaction. 
To be more precise, we pose the problem of a
one-dimensional model with
an interaction hamiltonian
\begin{equation}
H_{int} = \frac{e^2}{8\pi } \int dx dx'\;
\left( \Psi^{+}_L (x) \Psi_L (x) +
\Psi^{+}_R (x) \Psi_R (x) \right)
\frac{1}{|x - x'|} 
\left( \Psi^{+}_L (x') \Psi_L (x') +
\Psi^{+}_R (x') \Psi_R (x') \right)
\label{ham}
\end{equation}
where $\Psi_L (x)$ and $\Psi_R (x)$ are the electron field
operators for the left and the right branch of the linear
dispersion relation, respectively.
The RG method is a sensible approach to deal with this problem
since the $1/|x|$ interaction potential (as well as the
$\delta (x)$ potential) gives rise to a marginal four-fermion interaction.
The scaling dimension of the electron field
$\Psi (x)$ is $- 1/2$ , in length units. This means that the interaction
hamiltonian in (\ref{ham}) scales appropriately, with a 
dimensionless coupling constant $e^2$ (in units in which
$\hbar  = c = 1$), as the energy scale is 
reduced down to the Fermi level\cite{sh}.

However, the drawback in dealing with (\ref{ham}) is
that it contains a highly nonlocal operator, which makes unclear
the applicability of RG methods, usually devised to deal with a
set of local operators.
This problem can be circumvented by introducing a local
auxiliary field to propagate the Coulomb interaction. 
The hamiltonian can be written in the form
\begin{equation}
H  =  i v_F \int dx \; \left( \Psi^{+}_R (x) \partial_x \Psi_R (x)
 -  \Psi^{+}_L (x) \partial_x     \Psi_L (x)  \right)
 +   e \int dx \; \left( \Psi^{+}_L (x) \Psi_L (x) +
\Psi^{+}_R (x) \Psi_R (x) \right) \; \phi (x)
\label{ham2}
\end{equation}
where the $\phi (x)$ field propagates the interaction
\begin{equation}
i \langle T  \phi (x,t) \; \phi (x',t') \rangle = \frac{1}{4\pi}
\delta (t - t') \frac{1}{|x - x'|} \;.\;
\end{equation}

We may think of $\phi (x)$ as the scalar potential in 
three-dimensional quantum
electrodynamics.  However, the differences with that theory in
the present case are notorious since the propagation of $\phi
(x)$ is that of a genuine field in three spatial dimensions,
while the electrons are confined to one dimension. In general,
one may expect a better infrared behavior in the present model.
The propagator of the $\phi (x)$ field in momentum space can be
read from the relativistic expression\cite{landau}, 
after sending the speed
of light to infinity,
\begin{equation}
i \langle T  \phi (x,t) \; \phi (x',t') \rangle = \int \frac{dq
d\omega}{(2 \pi)^2} \int \frac{dq_y dq_z}{(2 \pi)^2}
\frac{\mbox{\Large $e^{i q (x - x')}$ }
\mbox{\Large $e^{-i \omega (t - t')}$ }
  }{ q^2 + q_y^2 + q_{z}^{2} - i \epsilon } 
\label{prop}
\end{equation}
The usual one-dimensional propagator 
$\sim  \log(|q|/\Lambda ) $ is recovered from (\ref{prop}) 
upon integration of the
dummy variables $q_y$ and $q_z$. We remark that the ultraviolet
cutoff $\Lambda $ for excitations along the $y$ and $z$
transverse directions is needed when projecting the
three-dimensional interaction down to the one-dimensional
system. 

The usefulness of the representation (\ref{prop}) can be
appreciated in the renormalization of the model at the one-loop
level. We study the scaling behavior of the irreducible
functions as the bandwidth cutoff $E_c$ is sent towards the
Fermi level, $E_c \rightarrow 0$. The self-energy to the
one-loop order is
\begin{equation}
i \Sigma (k,0) = i e^2 \int^{E_c}_{-E_c} \frac{ dp}{2\pi }
 \int^{+\infty}_{-\infty} \frac{ d\omega_p}{2\pi }
\frac{v_F (p + k)}{\omega_p^2 + v_F^2 (p + k)^2} \int \frac{dp_y
dp_z}{(2\pi)^2} \frac{1}{p^2 + p_y^2 + p_z^2} 
\label{pert}
\end{equation}
The limit $k \rightarrow 0$ has to be taken carefully in this
expression, by first combining the two denominators with the use
of Feynman parameters\cite{landau}. Finally we get
\begin{eqnarray}
i \Sigma (k,0) & = &  \frac{i}{4\pi} \frac{e^2}{v_F}
\int_{0}^{1} du \frac{1}{\sqrt{u}} \int_{-E_c}^{E_c}
\frac{dp}{2\pi} \int_{-\infty}^{\infty} \frac{d\omega_p}{2\pi}
\frac{v_F k} {\omega_p^2 + p^2 + v_F^2 k^2 u(1-u) }   \nonumber
  \\ &  \approx  & i \frac{e^2}{4\pi^2} k \; \log E_c  
\label{ren}
\end{eqnarray}
The term linear in $k$ in $\Sigma (k,0)$ represents a
renormalization of the Fermi velocity, which grows upon
integration of the high-energy modes. Obviously, there is no
correction linear in $\omega_k$ renormalizing the electron
wavefunction at the one-loop level. This is consistent with the
fact that the integration of high-energy modes at $\sim E_c$
does not renormalize the three-point vertex $\Gamma $ . 
We stress the difference of the logarithmic
renormalization of $v_F$ with respect to the usual {\em finite}
corrections due to a short-range interaction. The nontrivial
scaling of $v_F$ is a genuine effect of the long-range
Coulomb interaction, which also takes place in higher
dimensions\cite{ggv2,np}. 

Incidentally, the above computation exemplifies how
the Ward identity that ensures the integrability of the
Luttinger model does not hold in the present case. The Ward
identity is a relation between the electron Green function
$G(p,\omega_p)$ and the three-point vertex
$\Gamma (p, \omega_p ; k,\omega_k)$ at a given branch\cite{sol}.
For the right-handed modes, for instance, it is
\begin{equation}
\Gamma (p, \omega_p ; k, \omega_k) =
 \frac{G^{-1} (p, \omega_p) - G^{-1} (p-k, \omega_p - \omega_k)}
{\omega_k - v_F k} 
\label{wi}
\end{equation}
By focusing on the singular dependences on the bandwidth
cutoff $E_c$, one can check that (\ref{wi}) is already violated in our
model to first order in perturbation theory. Actually, the
dependence of the vertex $\Gamma $ on the variables $(k,\omega_k)$
of the external interaction line is 
\begin{equation}
\Gamma (p, \omega_p ; k, \omega_k) \approx 1 -
\frac{e^2}{4\pi^2} \frac{k \; \log \Lambda }{\omega_k - v_F k}
+  \ldots   
\end{equation}
We notice that the Ward identity would be satisfied if the
scaling could be implemented simultaneously in the transverse
ultraviolet cutoff $\Lambda $ and the bandwidth cutoff $E_c$,
that is, by taking $\Lambda = E_c$.
However, in a real system the scaling in $\Lambda $ gets
locked by the finite cross section of the wire, while only the
scaling in the longitudinal direction operated by $E_c$ is
allowed. In this respect, the gauge invariance of
quantum electrodynamics is broken by the
anisotropy of the electron system, as felt by the propagation of
the three-dimensional electromagnetic field.

The renormalization of $v_F$ at the one-loop level is not, in general, 
a sensible effect from the physical point of view, since the
propagator of the $\phi (x)$ field is drastically modified by
the quantum corrections. 
In what follows we implement a GW approximation in order to take
into account the dynamical screening due to plasmons.
The suitability
of this approximation in the study 
of one-dimensional systems has been recently shown in Ref.
\onlinecite{gw}.
The same approach has been also tested in the study of the crossover 
from Fermi liquid to Luttinger liquid behavior\cite{dicastro},
as well as in the study of singular interactions in dimension
$1< d \leq 2$ \cite{wen,mac}.

The self-energy $\Pi (k, \omega_k)$ of the $\phi (x) $ field is 
given at the one-loop level by the sum of particle-hole
diagrams with modes in the left and the right branch of the
dispersion relation. The number of contributions depends on the
number $n$ of different Dirac fermions, so that
\begin{equation}
i \Pi (k, \omega_k) = i n  \frac{e^2}{\pi} \frac{\tilde{v}_F k^2}
 {\tilde{v}_F^2
k^2 - \omega_k^2 }  
\label{pol}
\end{equation}
In the case of carbon
nanotubes we have $n = 4$, taking into account the two Fermi 
points and the spin degeneracy , but
it is also conceivable that in the process of renormalization
$n$ is given effectively by the number of subbands within the
energy cutoff $E_c $. In the spirit of the GW approximation,
we consider $\tilde{v}_F$ as a free parameter that has to match
the Fermi velocity in the fermion propagator after self-energy
corrections. We recall that the result
(\ref{pol}) turns out to be the exact polarization operator in
the model with short-range interactions, with an unrenormalized
Fermi velocity $\tilde{v}_F = v_F$ \cite{dl}.
In the present case, though, the violation of the Ward identity
(\ref{wi}) already signals that we do not have at work
the precise cancellation between
electron self-energy insertions and vertex corrections
characteristic of the Luttinger model. The relevant remnant is
actually the renormalization (\ref{ren}) of $v_F$,
and it can be taken into account self-consistently by replacing
$\tilde{v}_F$ in (\ref{pol}) by the renormalized value of the
Fermi velocity.

The particle-hole processes lead to a
modified propagator of the $\phi (x)$ field
\begin{equation}
i \langle \phi (k,\omega ) \; \phi (-k,-\omega ) \rangle =
 1 / \left( -\frac{2 \pi}{\log(|k|/\Lambda) } + \Pi (k,\omega ) \right)
\label{prop2}
\end{equation} 
The expression (\ref{prop2}) provides a sensible
approximation for the scalar propagator, as it incorporates
the effect of plasmons in the model.
Thus, our approach is that of using the scalar propagator
(\ref{prop2}) in the renormalization of the Fermi velocity and
the electron wavefunction. We compute the
electron self-energy by replacing the Coulomb potential by
the dressed interaction (\ref{prop2})
\begin{equation}
i \Sigma (k, i \omega_k) = i \frac{e^2}{2\pi } \int^{E_c}_{-E_c} 
\frac{dp}{2\pi }
 \int^{+\infty}_{-\infty} 
\frac{d\omega_p}{2\pi } \frac{1}{i (\omega_p + \omega_k) - v_F (p +
k) } \frac{\log(|p|/\Lambda) }{1 - n \frac{e^2}{2\pi^2 } 
 \frac{\tilde{v}_F p^2}{\tilde{v}_F^2
p^2 + \omega_p^2 } \log(|p|/\Lambda) } 
\label{selfe}
\end{equation}

Alternatively, one can think of the diagrammatics encoded in 
Eq. (\ref{selfe}) as the leading order in a $1/n$
expansion, in a model with $n$ different electron flavors.  
It can be shown that this approximation 
reproduces the exact anomalous dimension of the electron
field in the Luttinger model with conventional short-range
interaction\cite{equ}.
In our case, such approximation to the self-energy is also
justified since it takes into account, at each level in
perturbation theory, the most singular contribution at small
momentum transfer of the interaction. 
Due to the cancellation of fermion loops with more than 
two interaction vertices which
still takes place in the same way as in the Luttinger model,
the representation
(\ref{selfe}) for the self-energy only misses
the effects of vertex and electron self-energy corrections, which
may be incorporated consistently in the RG framework by
an appropriate scaling of the renormalized parameters.

The only contributions in (\ref{selfe}) depending on the
bandwidth cutoff are terms linear in $\omega_k$ and $k$. In
this respect, 
it is worth mentioning that, although the usual perturbative approach
gives rise to poles of the form $k^2/(\omega_k - v_F k)$ in the
self-energy\cite{dl}, these do not arise in the GW approximation. 
There is no infrared catastrophe at $\omega_k \approx v_F k$,
because of the correction in the slope of the plasmon dispersion
relation with respect to its bare value $v_F$. The
result that we get for the renormalized electron propagator is
\begin{eqnarray}
G^{-1}(k,\omega_k) & = &  Z^{-1}_{\Psi} \; (\omega_k - v_F
  k)  - \Sigma (k,\omega_k)  \nonumber   \\ 
& \approx &
Z^{-1}_{\Psi} \; (\omega_k - v_F  k) +
Z^{-1}_{\Psi} \; (\omega_k - v_F k) \frac{1}{n} \int^{E_c} 
 \frac{dp}{|p|} r^2
\frac{(1 - f(p))^2 }{2 \sqrt{f(p)} \left( 1 + r \sqrt{f(p)}\right)^2 }
                  \nonumber               \\ 
&  &  -
Z^{-1}_{\Psi} \; k \; \frac{e^2}{4\pi^2 }\int^{E_c} 
 \frac{dp}{|p|} 
   \frac{ f(p)^{3/2} + (4r/3 + r^3/3) f(p) + r^2 \sqrt{f(p)} + r/3}
  { f(p)^{3/2} \left( 1 + r \sqrt{f(p)} \right)^3  }
\label{gren}
\end{eqnarray}
where $f(p) \equiv 1 - n e^2 \log(|p|/\Lambda) /(2 \pi^2 \tilde{v}_F) \;, 
\; r \equiv \tilde{v}_F / v_F  \; $ and
$Z^{1/2}_{\Psi}$ is the scale of the
bare electron field compared to that of the cutoff-independent
electron field
\begin{equation}
\Psi_{bare}(E_c) = Z^{1/2}_{\Psi} \Psi \;.\;
\end{equation}

\begin{figure}

\par
\centering
\epsfbox[0 0 364 364]{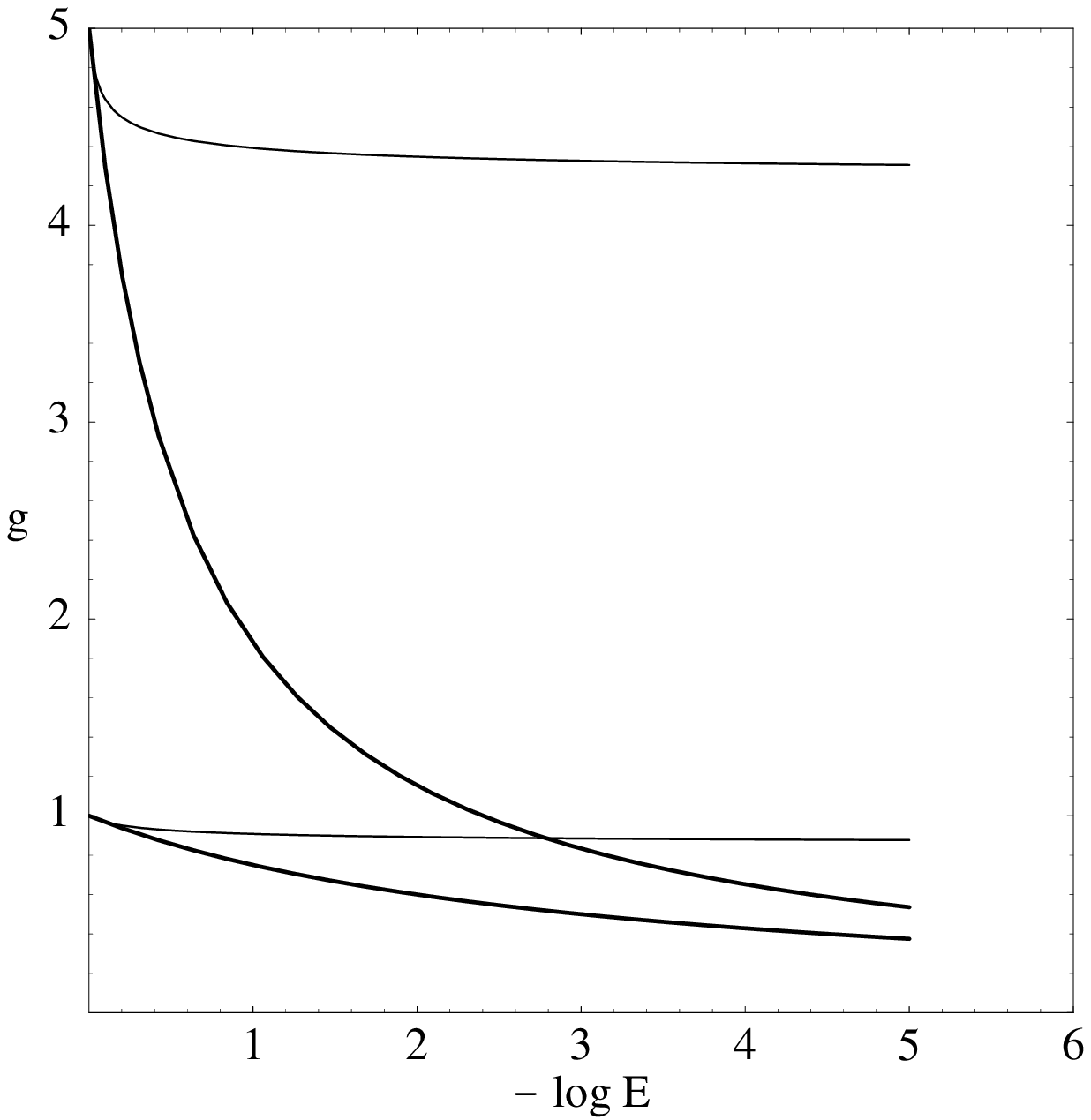}
\par

\vspace{2cm}
\caption{Flow of the effective coupling constant for different
bare values, computed in the large-$n$ solution (thick lines) and
in the model with only one band (thin lines).}
\label{two}
\end{figure}

In the RG approach, we
require the cutoff-independence of the renormalized Green
function, since this object leads to observable quantities in
the quantum theory. For this purpose,
the quantities $Z_{\Psi}$ and $v_F $ have to be promoted to
cutoff-dependent
effective parameters, that reflect the behavior of the quantum
theory as $E_c \rightarrow 0$ and more states are integrated out
from high-energy shells of the band. Regarding the problem of
self-consistency for the renormalized value $\tilde{v}_F$ of the
Fermi velocity, we find two possible solutions leading
to different physical pictures:

i) large-n limit solution.
In this limit we know that the expression (\ref{pol}) gives the
exact result for the polarization operator, since any vertex 
or self-energy correction makes any diagram subdominant from the point
of view of the $1/n$ expansion. Then the correct choice for 
$\tilde{v}_F$ has to be the fixed-point value of the Fermi velocity.
Self-consistency is therefore attained by requiring that the
solution of the scaling equation
\begin{equation}
E_c \frac{d}{dE_c } \: v_F (E_c) = - \frac{e^2}{4 \pi^2 }
\frac{ f(E_c)^{3/2} + (4r/3 + r^3/3) f(E_c) + r^2 \sqrt{f(E_c)} + r/3}
  { f(E_c)^{3/2} \left( 1 + r \sqrt{f(E_c)} \right)^3  }
\label{largen}
\end{equation}
matches the fixed-point  $\tilde{v}_F$ in the limit $E_c 
\rightarrow 0$. It can be checked, however, that any finite value
of $\tilde{v}_F$ in the right-hand-side of (\ref{largen}) does not
lead to a strong enough flow to reach $\tilde{v}_F$ at 
$E_c = 0$. The only self-consistent solution is found for 
$\tilde{v}_F = \infty$. For this value the right-hand-side of 
(\ref{largen}) has a finite limit, which  produces 
the asymptotic scaling of the Fermi velocity $v_F (E_c)
\sim - e^2 / (12 \pi^2) \:  \log (E_c )$.

ii) one-band solution.
If we stick to the picture in which we only pay attention
to the left and right linear branches of the dispersion 
relation, we have to assume that (\ref{pol}) only provides 
an approximate expression for the polarization operator.
The incomplete cancellation between electron self-energy and
vertex corrections to that object arises from the mismatch
between $E_c$ and the transverse cutoff $\Lambda $. 
As a consequence of that, the renormalized Fermi velocity in
(\ref{pol}) gets an effective dependence on the variable
$\log (E_c / \Lambda )$ and $\tilde{v}_F $ is to be taken 
as the scale dependent Fermi velocity, $\tilde{v}_F = v_F (E_c)$.
With regard
to the carbon nanotubes, this is consistent with the regime 
in which the scaling has progressed to distances larger than 
the radius of the nanotube. 
The RG flow equations then turn out to be 
\begin{eqnarray}
E_c \frac{d}{dE_c }\:  \log \: Z_{\Psi}(E_c)  & = & 
\frac{ \left( 1 - \sqrt{f(E_c)} \right)^2 }
    {8 \sqrt{f(E_c)}}     \label{zflow}  \\
E_c \frac{d}{dE_c } \: v_F (E_c)  & = &  - \frac{e^2}{4\pi^2}
    \frac{\sqrt{f(E_c)}  -  4/3  +  1/ \left(3 f(E_c)^{3/2} \right)}
    {\left( 1-f(E_c) \right)^2}  
\label{vflow}
\end{eqnarray}
Eq. (\ref{vflow}) is now the requirement of self-consistency, 
with $\tilde{v}_F = v_F (E_c)$. 
As mentioned before, the three-point
vertex only gets the cutoff dependence given by the wavefunction
renormalization in (\ref{zflow}). This means that the electron
charge is not renormalized at low energies in our field
theory framework. The behavior of the effective interaction is
therefore completely encoded in Eq. (\ref{vflow}), which can
be rewritten for the effective coupling constant $g \equiv
e^2 /(4\pi^2 v_F)$ in the form
\begin{equation}
E_c \frac{d}{dE_c } g(E_c) = \frac{1}{64(\log \: E_c)^2} \left(
 \sqrt{f(E_c)} - \frac{4}{3} + \frac{1}{3f(E_c)^{3/2}} \right)
\label{gflow}
\end{equation}
The right-hand-side of Eq. (\ref{gflow}) vanishes as
$e^2 /(4\pi^2 v_F) \rightarrow 0$ and it
could still admit a solution of the form
$g \sim - g_0 / \log \: E_c $, but this is not realized 
in the present regime as the
equation $8 g_0 = \sqrt{1+g_0} - 4/3 + 1/(3\sqrt{(1+g_0)^3})$
does not have any real solution. The flow of $g (E_c)$ 
given by (\ref{gflow})
quickly approaches some fixed-point value, after
which it becomes little sensitive to further scaling 
in the infrared\cite{prev}.
A plot of the flow for different values of the 
bare coupling constant at $E_c = \Lambda $
is given in Fig. \ref{two}, which includes for comparison the
scaling behavior of the large-$n$ solution.

The existence of two different regimes in the model
is not surprising, since
the dependence on the number $n$ of subbands is the way in which
the system keeps memory of the finite transverse dimension
in the carbon nanotube. A nanotube of very large radius, for
instance, leads to a picture in which a large number of subbands
are stacked above and below the linear branches in Fig. \ref{one}.
This system falls into the description of the model with a very
large $n$ value. On the other hand, there is always a sufficiently 
small value of $E_c$ for which all the subbands, but those contributing to
the gapless part of the spectrum, become higher in energy than 
the RG cutoff. At that point, one proceeds paying attention to the
linear branches crossing at the Fermi points alone, what leads
to the one-band regime of the model.

>From the point of view of the real space, the above transition
in the number $n$ of subbands represents the crossover from the
two-dimensional regime to the effective one-dimensional
description of a nanotube of large radius. Actually, in the 
limit of a very large $n$ 
we recover the scaling behavior of the
effective coupling  constant in the graphite layer, which flows
towards the free fixed-point. This scaling of the coupling
constant is actually what makes consistent the linear
quasiparticle decay rate recently measured\cite{exp}
with the metallic properties of graphite\cite{prl,prb}.
In a carbon nanotube,
the scaling appropriate for the graphite regime has to be followed
up to a distance scale of the order of the diameter of the nanotube.
The value of the coupling constant at that scale is  
what dictates the bare value for the one-dimensional regime
ii). The one-dimensional effective field theory contains the
explicit dependence on the ultraviolet cutoff $\Lambda $, that
is actually needed to fix the conditions at the crossover
between the two regimes.

The issue 
of the renormalization of $v_F$ we have discussed is important,
since it implies a reduction in the strength
of additional interactions in the system, whether short or
long-range, as the effective couplings are all given by the
couplings in the interaction hamiltonian divided by $v_F$.  
We stress that this renormalization of the Fermi velocity 
is the relevant effect for a phenomenology at realistic energies, because
the short-range interactions like umklapp or backscattering are nominally
tiny and their RG flow only becomes appreciable at 
extremely small energies.

Incidentally, such a renormalization of $v_F$ affecting every
effective interaction in the model
may also be present in small chains, contributing to explain
the insulator-metal transition by the effect of the Coulomb
interaction observed in the exact diagonalization of finite
rings\cite{poil}. The study of low-dimensional systems carried out
in Ref. \onlinecite{saw}
also indicates a reduction of the electron correlations similar to
our findings in the RG framework.

The reason for the distinctive
behavior of the Coulomb interaction in the fullerene tubules is
the preservation of the chiral invariance in such systems. This
comes from the different symmetry properties that characterize
the left and the right modes at each Fermi point, leading to the
selection rule that forbids their hybridization by the
interaction. Thus, although it has been proposed that in a
generic one-dimensional electron system the $2 k_F$ and higher
harmonics may lead to strong correlation effects in the presence
of the Coulomb interaction\cite{schulz}, the absence of such hybridization
makes the carbon nanotubes fall into the more conventional
Luttinger liquid regime.
As stated above, the wavefunction renormalization (\ref{zflow})
is just the generalization for the long-range interaction of 
the expression giving the anomalous dimension of the electron
field in the Luttinger model. The important point is the 
existence of a stable fixed-point for the effective coupling
constant in the infrared. All the properties of the
Luttinger liquid universality class are then expected to hold
in the metallic carbon nanotubes, with a sensible renormalization
of the effective interaction strength depending on the radius of the
tubule.

We thank C. Castellani for useful discussions. This work has been
partially supported by a CICYT-INFN exchange program and by the 
spanish Ministerio de Educaci\'on y Cultura Grant No. PB96-0875.

\newpage

\newpage

\begin{references}
\bibitem{equ}
It can be checked that the right-hand-side of Eq. (\ref{zflow})
matches the critical exponent of the electron propagator 
in the Luttinger model,
after replacement of $- e^2 \log (E_c /\Lambda )$ by a constant
interaction $g_2 = g_4$.
\end{references}
\end{document}